\documentclass[
 aip,
jmp,
 amsmath,amssymb,
preprint,%
author-year,%
]{revtex4-1}

\usepackage{graphicx}
\usepackage{dcolumn}
\usepackage{bm}
\usepackage[mathlines]{lineno}
\usepackage[utf8]{inputenc}
\usepackage[T1]{fontenc}
\usepackage{mathptmx}
\usepackage{etoolbox}

\makeatletter
\def\@email#1#2{%
 \endgroup
 \patchcmd{\titleblock@produce}
  {\frontmatter@RRAPformat}
  {\frontmatter@RRAPformat{\produce@RRAP{*#1\href{mailto:#2}{#2}}}\frontmatter@RRAPformat}
  {}{}
}%
\makeatother
\begin{document}

\preprint{arXiv:2201.02094v2}

\title[Acoustic velocity at high pressures and temperatures]{A high pressure, high temperature gas medium apparatus to measure acoustic velocities during deformation of rock}
\author{C. Harbord}
 \email{c.harbord@ucl.ac.uk}
\author{N. Brantut}%
\author{E.C. David}
\author{T.M. Mitchell}
\affiliation{ 
Department of Earth Sciences, University College London, Gower Place,  London, WC1E 6BT, UK
}%

\date{\today}

\begin{abstract}
A new set-up to measure acoustic wave velocities through deforming rock samples at high pressures (up to 1000 MPa), temperatures (up to 700$^\circ$C) and differential stress (up to 1500 MPa) has been developed in a recently refurbished gas medium triaxial deformation apparatus.
The conditions span a wide range of geological environments, and allow us to accurately measure differential stress and strains at conditions which are typically only accessible in solid medium apparatus.
Calibrations of our newly constructed internal furnace up to 1000 MPa confining pressure and temperatures of up to 400$^\circ$C demonstrate that the hot zone is displaced downwards with increasing confining pressure, resulting in temperature gradients that are minimised by adequately adjusting the sample position.
Ultrasonic velocity measurements are conducted in the direction of compression by the pulse-transmission method.
Arrival times are corrected for delays resulting from the geometry of the sample assembly and high-precision relative measurements are obtained by cross-correlation.
Delays for waves reflected at the interface between the loading piston and sample are nearly linearly dependent on differential applied load due to the load dependence of interface stiffness.
Measurements of such delays can be used to infer sample load internally.
We illustrate the working of the apparatus by conducting experiments on limestone at 200 MPa confining pressure and room temperature and 400$^\circ$C.
Ultrasonic data clearly show that deformation is dominated by microcracking at low temperature and by intracrystalline plasticity at high temperature.

\end{abstract}

\maketitle

\section{\label{sec:level1}Introduction}

The study of rock deformation at lithospheric conditions is technically challenging since it requires generating high pressures, temperatures and differential stresses.
To simulate shallow crustal conditions  (i.e., less than 20 km), corresponding to confining pressures of up to 400 MPa and temperatures of up to 200$^\circ$C, triaxial oil medium apparatus are used.
They can accommodate large samples (20-100 mm in diameter), and are also typically furnished with pore fluid pressure systems (up to 200 MPa) to allow independent control of confining and pore pressures.
Due to their ease of operation, oil-medium triaxial systems are the most widely used apparatus in rock deformation, although they are generally limited to investigations of brittle rock failure only.

\begin{figure}
\includegraphics[scale=1]{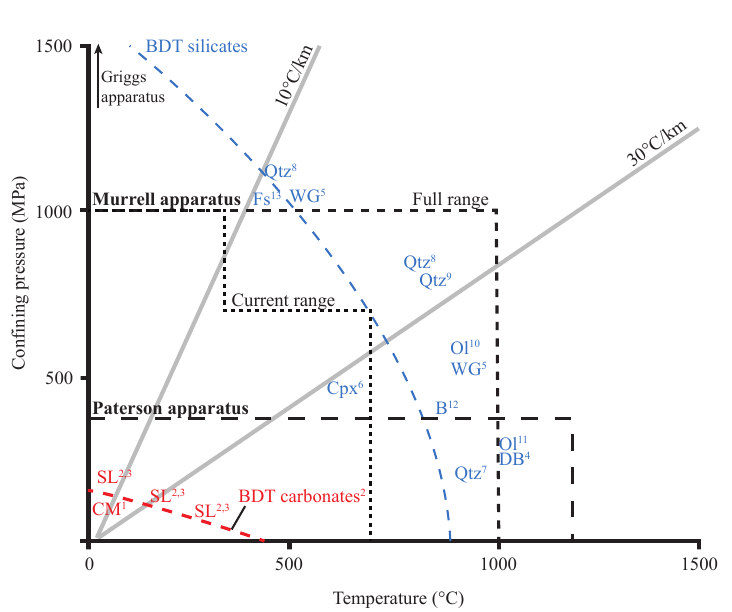}
\caption{\label{Fig:range}Design and current operating range of the Murrell triaxial apparatus compared to the approximate onset of the 'dry' brittle-ductile transition in rocks (BDT in figure), modified after \protect \citep{Murrell1990b}. The range of Paterson apparatus is also shown for comparison \citep{Paterson1970}, and we also note that the displayed range of conditions can be easily achieved using Griggs' apparatus \protect\citep{Griggs1967}. Legend: CM- Carrara marble (1. \protect \cite{Fredrich1989}), SL- Solnhofen limestone (2. \protect \cite{Heard1960}, 3. \protect \cite{Rutter1974}), DB- Maryland diabase (4. \protect\cite{Caristan1982}), Cpx- Clinopyroxenite (6. \protect \cite{Kirby1984}),  Qtz- Quartzite (7. \protect \cite{Kanaya2018}, 8. \protect \cite{Hirth1994}, 9. \protect \cite{Heard1968}), Ol- olivine (10. \protect\cite{Raleigh1968}, 11. \protect\cite{Chopra1981}), B- basalt (12. \protect\cite{Violay2012}), Fs- feldspar (\protect\cite{Tullis1992}). The grey lines show typical geothermal gradients of subduction zones (10 $^\circ$C/km) and mountain belts (30$^\circ$C/km), where rocks are actively deformed in the earth's lithosphere.}
\end{figure}

More extreme conditions (i.e., greater than pressures of 400 MPa, temperatures of 200$^\circ$C and differential stresses 500 MPa) require the use of more specialised deformation apparatus, where confinement is either provided by a solid medium in a piston-cylinder configuration \citep[e.g.,][and figure \ref{Fig:range}]{Griggs1967}, or an inert gas in a triaxial configuration \citep[e.g.,][and figure \ref{Fig:range}]{Paterson1970}.
Commonly used "Griggs" solid medium apparatus can deform rock samples, up to confining pressures of 3500 MPa, differential stresses of 3000 MPa, and temperatures of 1500$^\circ$C, representative of the middle crust to upper mantle (i.e., up to 150 km burial depth, figure \ref{Fig:range}).
"Paterson" apparatus are the most commonly used gas medium triaxial deformation systems, and are currently used up to confining pressures of 500 MPa, temperatures of 1500$^\circ$C and differential stresses of 1000 MPa (figure \ref{Fig:range}).
In the past, gas apparatus have been used up to confining pressures of 1000 MPa \citep{Heard1960,Paterson1970}.
However, a pressure vessel failure at 500 MPa led Paterson to restrict routine confining pressure conditions to 300 MPa \citep{Paterson1970}. 

"Griggs" and "Paterson" apparatus are used to study different deformation regimes.
Rock samples deformed in a "Griggs" apparatus are relatively small (of 6.3 mm in diameter and 12 mm in length).
Axial load and confining load are measured externally, which results in inaccuracies to the determination of the internal stress state \citep{Holyoke2010}. 
Therefore, solid medium apparatus are suited to deforming materials in high stress regimes, and careful calibrations are required to obtain reliable strength measurements \citep[e.g., ][]{Gleason1995,Holyoke2010}.
In a "Paterson" apparatus, samples are larger (10--15 mm in diameter and about 20 mm in length) and differential stress is measured internally.
These factors result in a considerably better accuracy in measuring differential stress and strain, making gas apparatus ideal to study material deformation in the high temperature, low stress regimes.
However, their limited differential stress capacity (up to 1000 MPa) makes them less suitable for work on low temperature plasticity.
Recent work by \citet{Burdette2021} has started to bridge the gap between the traditional capabilities of solid medium "Griggs" and gas medium "Paterson" apparatus, by significantly improving load resolution measurements in a modified "Griggs" system equipped with a laser interferometer.

Generally, rock deformation experiments are complemented by post-mortem microstructural investigations.
As a result of unloading, decompression and cooling, post mortem microstructural investigations are prone to artefacts, it is therefore desirable to measure \textit{in situ} physical properties during deformation.
\textit{In situ} measurements are easiest in oil medium triaxial apparatus, where large sample dimensions permit the use of a wide range of sensors.
At more extreme test conditions, measurements of physical properties are significantly more difficult to obtain.
Due to the limited sample size and access, Griggs apparatus are the most challenging apparatus to instrument, and studies have been limited to acoustic emission \citep{Blacic1977,Ghaffari2020} and axial P-wave velocity \citep{Moarefvand2021}.
In gas apparatus, \textit{in situ} measurements during deformation have included volume changes \citep{Fischer1989}, permeability \citep{Fischer1992}, acoustic emission \citep{Burlini2007}, electrical conductivity \citep{Ferri2009}, and high frequency displacement \citep{Hayward2016}.
Other \textit{in situ} measurements at hydrostatic conditions in gas-apparatus include ultrasonic wavespeeds \citep[e.g., ][]{Christensen1979} and torsional attenuation \citep[e.g., ][]{Jackson1984}.

Here we document modifications that we have made to a recently refurbished high pressure vessel, the "Murrell"  \citep[after Professor Stanley Murrell, 1929-2004, see ][]{Edmond1973,Ismail1974,Jones1989,Murrell1990a}.
The apparatus is capable of deforming rock up to confining pressures of 1000 MPa, temperatures of 1000$^\circ$C and differential stresses of 1500 MPa.
First, we describe the mechanical and electrical components of our gas apparatus.
Results from calibration of our internal furnace at a range of confining pressures (200--1000 MPa) and temperatures (100--400 $^\circ$C) are reported.
Following this, we document our recently commissioned ultrasonic velocity measurement system.
We demonstrate how this new acoustic data can be used to obtain information about deformation mechanisms during experiments using limestone at room temperature and 400$^\circ$C.
We find that a non-negligible transmitted wave delay occurs as a result of interfacial compliance, and suggest appropriate correction methods.
We also show that the delay associated with interfacial compliance can be used to obtain internal load using reflected wave measurements.

\section{\label{sec:level1}Apparatus description}
\subsection{\label{sec:level2}Pressure vessel}

\begin{figure}
\includegraphics{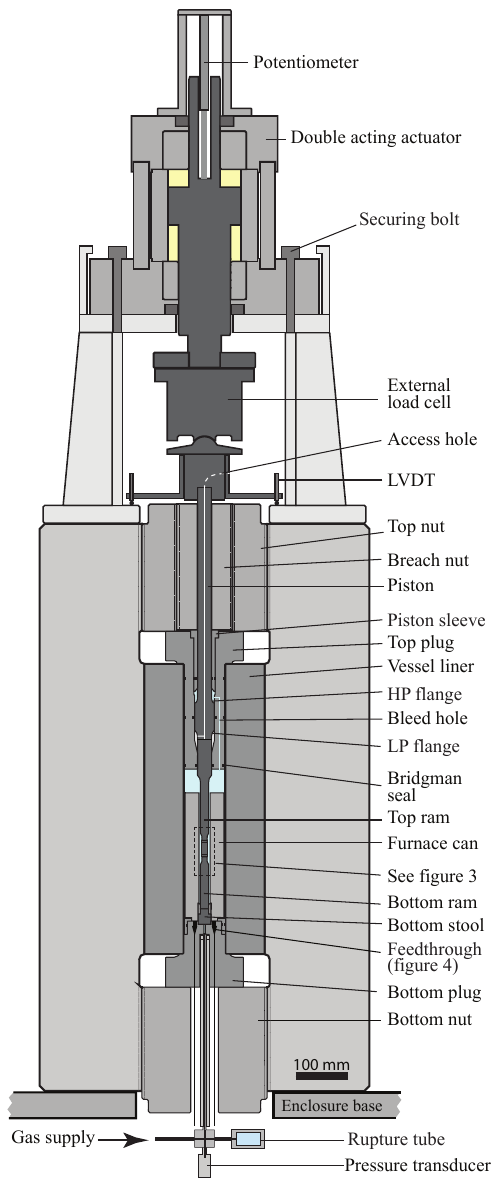}
\caption{\label{Fig:apparatus}Annotated scale sectional view of the Murrell deformation apparatus. 
Light blue shading indicates the location of the confining argon gas, and yellow shading indicates the location of hydraulic oil. 
LP flange is low pressure flange and HP flange refers to the high pressure flange. 
Further details of sealing arrangement is shown in figure \ref{Fig:bridgman}, sample assembly in figure \ref{Fig:sample} and feedthroughs in figure \ref{Fig:feedthrough}. }
\end{figure}

The pressure vessel, of a composite construction, is made of vacuum forge remelted Hecla tool steel (Figure \ref{Fig:apparatus}a).
The vessel has a maximum design confining pressure of 1400 MPa, a working confining pressure of 1000 MPa and a maximum working temperature of 1000$^\circ$C, using inert gas as a confining medium. 
Hecla 180 alloy (table \ref{Tab:composition}) is used for the vessel body (figure \ref{Fig:apparatus}) with a tensile strength of 1240 MPa (OD 635mm, ID= 220.05 mm) and is lined with Hecla 174 (figure \ref{Fig:apparatus} and table \ref{Tab:composition}) with a tensile strength of 1420 MPa (220.95mm OD, 80 mm ID).
The liner was inserted into the external vessel using shrink fitting.
An interference fit between the external vessel and liner results in a static compressive hoop stress at the outer radius of the liner. 
This equates to $\approx$400 MPa at the internal diameter of the vessel liner, and thus tangential stress in the vessel only becomes tensile when the internal confining pressure exceeds 400 MPa.

\begin{table}
\begin{tabular}{c|cccccccccc}
					&	C		&	Si		&	S			&	P			&	Mn	&	Cr		&	Ni		&	Mo	&	V 		& Fe	 	\\ 	\hline
Hecla 180 		&	.40	&	.27	&	.012		&	.016		&	.70	&	1.22	&	3.10	&	0.46	&	.16	& bal.	\\
Hecla 174		&	.42	&	1.18	&	.010		&	.016		&	.34	&	5.12	&	-		&	1.30	&	.95 	& bal.	\\	
\end{tabular}
\caption{\label{Tab:composition}Composition of the two types of steel constituting the pressure vessel.}
\end{table}

\subsection{\label{sec:level2}Axial piston and top plug}
Load is applied to samples through the top plug, using a piston with a double step in diameter to provide compensation for the effect of confining pressure (figure \ref{Fig:apparatus}).
The piston is made from D6 tool steel (58-60 Rc), with a length of 489 mm, a rod diameter of 27 mm and a piston section of 38 mm in diameter.
The piston section is flanged into the rod with two shoulder fillets.
The upper flange (HP flange, figure \ref{Fig:apparatus}) is elliptical in profile and forms part of the high pressure section of the compensation system, and is in a tensile stress condition.
The lower flange (LP flange, figure \ref{Fig:apparatus}) is circular in profile and forms part of the low pressure compensation chamber, and is subjected to compressive loads.
The upper flange of the piston was originally designed with a shoulder fillet of circular section, however after several catastrophic tensile failures of the piston, it was changed to an elliptical section \citep{Jones1989}, due to its favourable reduction of stress concentrations \citep{Peterson1974}.
The piston has a central bore of 4 mm diameter, allowing access for pore fluid or a range of sensors.
Its base is machined with an M12 thread to allow interchange of the rams.

\begin{figure}
\includegraphics[scale=1]{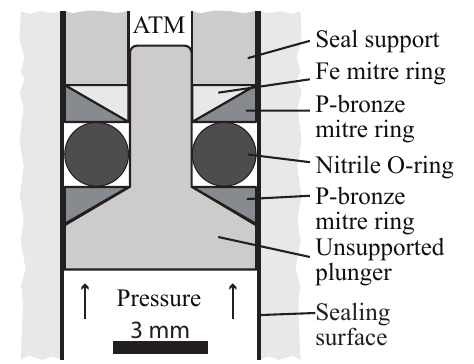}
\caption{\label{Fig:bridgman}Detailed sectional scale sketch of annular Bridgman seals following \protect\citet{Griggs1936} and \citet{Holloway1971a}, Fe = mild steel and P-bronze = phosphor bronze. See text for full description of sealing configuration.}
\end{figure}

Sealing of the piston and pressure autocompensation is achieved using 6 sets of annular Bridgman seals \citep[][ figure \ref{Fig:bridgman}]{Bridgman1914, Griggs1936, Holloway1971a, Tullis1986}, which are combined with a network of bores to distribute or vent the confining gas.
Bridgman seals in this configuration follow the descriptions of \citet{Griggs1936}, and comprise a stack of 3 mitre rings (two phosphor bronze and a single mild steel ring), which are chamfered at 30$^\circ$ and sandwich a single nitrile O-ring (figure \ref{Fig:bridgman}).
Pressure is transmitted onto the mitre rings and O-ring using a hardened D2 steel plunger (58-60 Rc), which are supported by the seal housing.
The seal arrangement is symmetric about a fin extending from the seal support backing and between the external and internal seal stack.
Nitrile O-rings with a shore hardness of 70 are used, which swell due to their high argon solubility.
To reduce seal friction seals are also lubricated with Molykote 55 O-ring grease, which also aids the swelling of the O-ring material.
The current sealing arrangement has been used up to a confining pressure of 1000 MPa without leakage, although considerable wear and damage is observed when using the seals above 600 MPa confining pressure.

\subsection{\label{sec:level2}Sample assembly}

\begin{figure}
\includegraphics{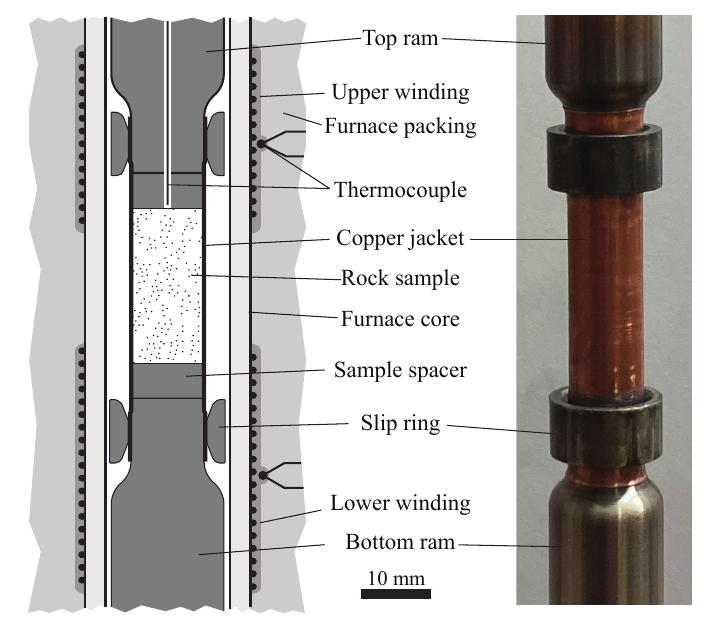}
\caption{\label{Fig:sample}Scale section view of sample assembly and furnace core, showing the key components of the configuration, supplemented by a picture of the sample assembly before insertion into the apparatus.}
\end{figure}

The sample assembly comprises a top and bottom ram, two sample spacers, two swaging rings, an annealed copper jacket and a rock sample (figure \ref{Fig:sample}). 
The top ram attaches to the base of the piston using an M12 stud, and the upper portion of the top ram forms a high pressure sealing surface with the lowermost Bridgman seals in the top plug (figure \ref{Fig:apparatus}).
Ram diameter is reduced using 3 large radius blends to the final 10 mm sample diameter. 
Sealing of the jacket is achieved by swaging tapered steel or nickel alloy rings over the ends of the jacket and the rams, where ram diameter increases from 10 to 10.4 mm, similarly to the technique used by \citet{Handin1953} and \citet{Heard1960}, see figure \ref{Fig:sample}.
To avoid stretching and work hardening of the jacket during swaging, which can cause jacket puncture, a light polish and a small quantity of MoS$_2$ grease is applied to the jacket surface before swaging.

A range of ram configurations is used depending on the experimental requirements.
For temperatures less than 200$^\circ$C rams are fabricated from D2 tool steel (58-60 Rc), which can be used up to the maximum working pressure of the vessel, and a differential stress of 1500 MPa.
For temperatures up to 700$^\circ$C the rams are fabricated from Inconel 718 (36-38 Rc), and can be used up to 1000 MPa differential stress and 630 MPa confining pressure.
In order to provide access for pore fluids or thermocouples a hollow top ram can be used.
In principle other modifications are possible to extend the range of conditions, or to integrate an internal load cell.

\subsection{\label{sec:level2}Bottom plug, bottom stool and feedthroughs}

\begin{figure}
\includegraphics{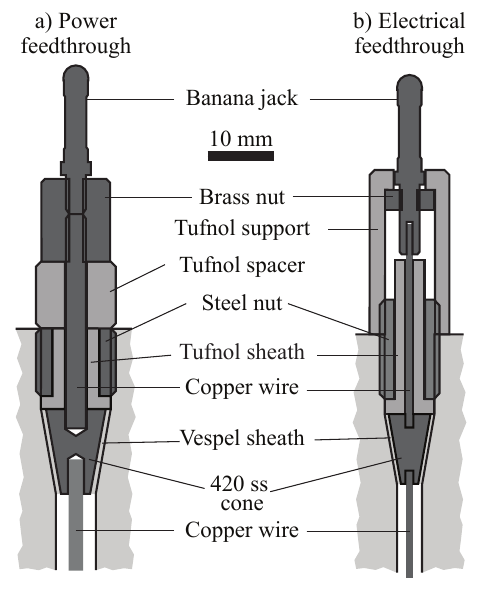}
\caption{\label{Fig:feedthrough}Detailed sectional scale sketch of feedthroughs in the bottom plug. Inset a) shows a power feedthrough, which is of large diameter for high currents. Inset b) shows a smaller diameter feedthrough used for electrical connections.}
\end{figure}

The bottom plug forms the lower closure of the pressure vessel, and is fabricated from 50 Rc Hecla 174 steel (table \ref{Tab:composition}).
A single annular Bridgman seal is used to seal the external diameter of the bottom plug (figure \ref{Fig:apparatus}).
The bottom plug is equipped with 12 Bridgman type feed through connections \citep{Bridgman1914}.
There are three 7 mm diameter high current power feedthroughs (figure \ref{Fig:feedthrough}a), and nine 5 mm diameter electrical feedthroughs (figure \ref{Fig:feedthrough}b).
Each feed through hole is machined with a 20$^\circ$ conical section, which forms the sealing surface against the feedthrough cores.
Feedthroughs are isolated from the bottom plug using Vespel sheaths on the 20$^\circ$ sealing surface, and Carp brand Tufnol on the high pressure side.
Historically, pyrophyllite was used to isolate feedthroughs, but it was found to be unreliable due to its poor machinability, high porosity and brittleness after fabrication.
PEEK plastic was also recently trialled, however it creeped severely, shorting the feedthrough connection.

The nine smaller holes are used for electrical connections (thermocouples or to pulse a piezo ceramic) and comprise a 420 stainless steel cone with two 1.2 mm diameter blind holes (figure \ref{Fig:feedthrough}b).
A short section of tinned copper wire is brazed into either end of the cones, six of which are connected to a chromel or alumel wire, acting as leads for K-type thermocouples.
The remaining three small feedthroughs can be used for a variety of purposes.
The three larger 7 mm diameter 420 stainless steel cones are used to supply power to the furnace.
The cones are constructed with two blind holes into which a section of copper rod is brazed in place.
Each feedthrough wire is terminated inside a copper banana jack which fits into the base receptacle of the furnace.

Argon gas is supplied to the vessel through a central bore drilled along the axis of the bottom plug, entering the vessel at the base of the bottom stool (figure \ref{Fig:apparatus}a).
The bottom stool acts as a receptacle for the bottom ram, and rests in a 6$\times$25 mm diameter recess machined into the centre of the bottom plug (figure \ref{Fig:PZT_detail}b).
A tapered ring attached to the bottom stool acts as a centering guide for the sample assembly which is inserted from above. 

\subsection{\label{sec:level2}Pressure generation and decompression}
 
Gas is compressed and decompressed using a computer controlled 4-stage hydraulically driven gas intensification system, manufactured by Sovereign Pressure Products Ltd, UK.
Control and monitoring of the system is performed by use of a human machine interface or a LabView program.
Before pressurisation, it is standard practice to perform two purge cycles to flush the system of atmospheric air.
A pressure of 600 MPa can be reached within 10-20 minutes, although pressurisation is performed in stages over a period of an hour to limit sample damage.
Decompression is also computer controlled using a sequence of hydraulically and pneumatically actuated flow restrictors. 
It is also performed in several steps, to avoid sample damage and limit condensation of water vapour along pipe fittings.

\subsection{\label{sec:level2}Actuator, load and displacement measurements}

A double acting servo hydraulic actuator attached to the top of the pressure vessel can apply loads of up to 200 kN (i.e., a differential stress of 2550 MPa on 10 mm diameter sample).
The back of the actuator is fitted with a large range potentiometer, which is used for coarse control of the actuator when not in contact with the piston.
Load from the actuator is transmitted through a 20 tonne load cell (Eliott brothers, SC-50000) onto a hemispherical seat and a head block (figure \ref{Fig:apparatus}).
Fine displacement is measured using two removable LVDT's attached to the head block (Sangamo, model GT5000-L25, LVDT figure \ref{Fig:apparatus}), which are used to control displacement during sample loading.

\subsection{\label{sec:level2}Furnace assembly and calibration}

The internal furnace is housed in a stainless steel sleeve (80 mm OD, 74 mm ID, 230 mm length), and follows the design of previous workers \citep{Paterson1970,Holloway1971a,Paterson1990}.
Two windings of Kanthal A1 wire (Al5.8Cr22Fe72.2) are positioned along the alumina furnace core (outer diameter of 23.4 mm, inner diameter of 17.4 mm, and 126 mm in length), and are wired in series (figure \ref{Fig:apparatus}c).
Windings are potted and affixed to the core using glass cement (Glassbond Sauereisen 78/3), and the core is held in place using Macor ceramic spacers.
The remaining volume between the core and the can is packed using a combination of alumina paper (Zircar ceramics APA 2) and fibre (Zircar ceramics Alumina Bulk Fiber). 
Basal female banana jack receptacles, used for electrical connections, are held in place using a Carp brand Tufnol plate, with either ends of the internal assembly secured using two brass plates.

Two K-type thermocouples are used to monitor the upper and lower winding temperature, and are vertically centred with respect to the individual windings.
A third "safety" thermocouple is positioned next to the stainless steel sleeve to monitor the vessel wall temperature. 
In order to avoid retempering the vessel liner steel by applying excessive temperatures, the furnace power is cut if the temperature measured by the safety thermocouple exceeds 400$^\circ$C.
All three thermocouple temperatures are monitored using an Eurotherm NanoDac controller.
Power to the windings is controlled using two Eurotherm Epack 1PH digital silicon controlled rectifiers wired in series with two step down transformers.
Both the NanoDac and E-pack units are controlled and logged using LabView and are calibrated for use with constant power.	

\begin{figure*}
\includegraphics{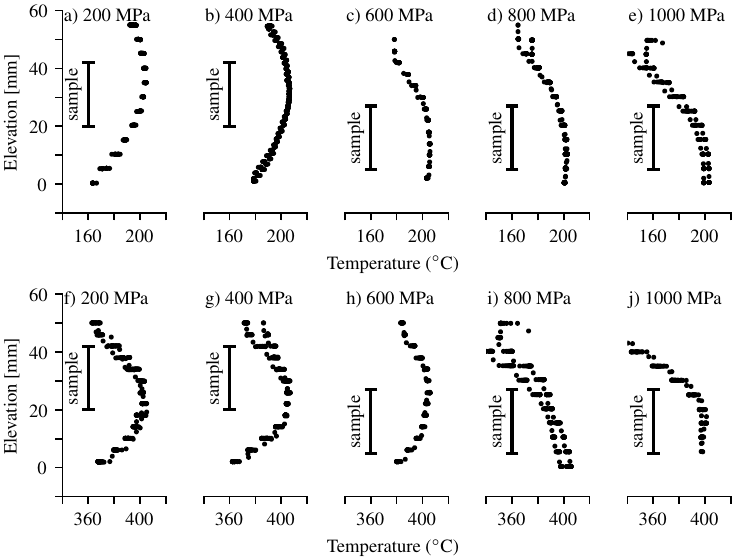}
\caption{\label{Fig:furnace}Calibration of the internal furnace at a range of confining pressures and temperatures, elevation is in mm and is relative to the top of the bottom ram. Insets a) to e) show calibrations at a target temperature of 200$^\circ$C, and 200 MPa confining pressure intervals up to 1000 MPa. Insets f) to j) show calibrations at a target temperature of 400$^\circ$C and at the same pressures as the 200$^\circ$C calibrations. The vertical bars show the positions that rock samples are used at each calibration condition.}
\end{figure*}
 
The power input to the furnace coils must be calibrated to produce target temperatures at the sample location at a range of pressure conditions.
The calibration was performed by measuring temperature profiles across the sample location using a K-type thermocouple inserted in a hollow alumina sample.
The thermocouple was attached to a miniature linear actuator (model: Actuonix L16P), with position measured using feedback from an onboard potentiometer.
The actuator is controlled from LabView using the libraries provided by Actuonix (https://www.actuonix.com/category-s/1930.htm), and profiling is repeated three times at each calibration condition (see figure \ref{Fig:furnace}).

Calibrations have been performed at 200 MPa intervals in confining pressure up to 1000 MPa, and at 100$^\circ$C temperature intervals up to 400$^\circ$C.
During the calibration, the thermocouple position was stepped at 2--3 mm intervals, in the range 0--60 mm above the top of the bottom stool (see figure \ref{Fig:sample} and \ref{Fig:furnace}).
The position of the thermocouple was held at each position for at least 30s, to capture temporal variations in temperature and allow the thermocouple tip to equilibrate with the ambient temperature.
The scatter at each point on the calibration plot reflects temperature fluctuations during hold periods, likely resulting from argon convection (figure \ref{Fig:furnace}).

Temperature profiles demonstrate that the elevation of the hot zone decreases with increasing pressure (contrast figure \ref{Fig:furnace}a and d).
We obtain temperatures within about 10$^\circ$C of the mean in the 150--200$^\circ$C range and 15$^\circ$C in the 300--400$^\circ$C range.
Increases in pressure tend to improve the temperature profile.
\citet{Paterson1970} previously observed a downward displacement of hot zone with increasing pressure, which can be explained by the increased density of argon at high pressure.
The temperature dependence of argon density is lower at higher pressure \citep{Gosman1969}, and therefore the buoyancy potential which drives convection is smaller. 
As a result, convective heat loss is suppressed,  resulting in a downwards displacement of the hot zone.
Based on the observed change in hot-zone elevation we introduced an extended bottom ram, with 15 mm extra length to position samples more favourably for lower pressure experiments.

\subsection{\label{sec:level2}Safety}

In order to reduce the risk posed by a catastrophic gas leak, several safety measures are employed.
The entire vessel and pressure generation system is enclosed in an interlocked steel cage (25 mm in thickness), which is lined with wood to absorb projectile energy. 
The base of the enclosure made from a thin plate steel, with the bottom plug exposed to the ground, which would direct any blast energy downwards.
Bolts are used to secure individual enclosure components together, with slots machined in the support structure to allow the enclosure to rise during a catastrophic failure.
Full rig operation can be performed remotely, however when the user is working in the laboratory ear defenders are worn at all times.
The vessel and seals are typically checked every 3 months for signs of significant wear or corrosion, and the bottom plug and nut is removed on an annual basis to perform a full vessel inspection.
To prevent the vessel and fittings being overpressured, a 1120 MPa rated rupture tube is installed at the base of the apparatus (figure \ref{Fig:apparatus}).

\subsection{\label{sec:level2}Data reduction}
Raw mechanical data must be corrected in order to obtain the true stress-strain response.
The absolute value of top plug seal friction is subtracted from load cell measurements before sample contact is made.
The seal friction is typically around 4.5 kN at 200 MPa, increasing to around 45 kN at 800 MPa, is not sensitive to internal temperature, and is stable with displacement (see figure S1).
As seal friction is dependent on confining pressure, an additional correction for seal friction is applied to the data after the hitpoint to account for changes in confining pressure.
A further contribution for the fraction of load supported by the copper jacket is computed from the creep equation of copper given by \citet{Frost1982}.
We do not take into account strain hardening of the jacket material.
Load point displacement is corrected for the loading stiffness which contributes to the recorded LVDT displacement.
Loading stiffness obtained by deformation of a steel blank was determined to be 68.9 kN mm$^{-1}$ for the D2 sample assembly and 65 kN mm$^{-1}$ for the Inconel 718 sample assembly.

\section{\label{sec:level1}Ultrasonic velocity measurements}


\subsection{\label{sec:level2}Sensor array}
\begin{figure}
\includegraphics{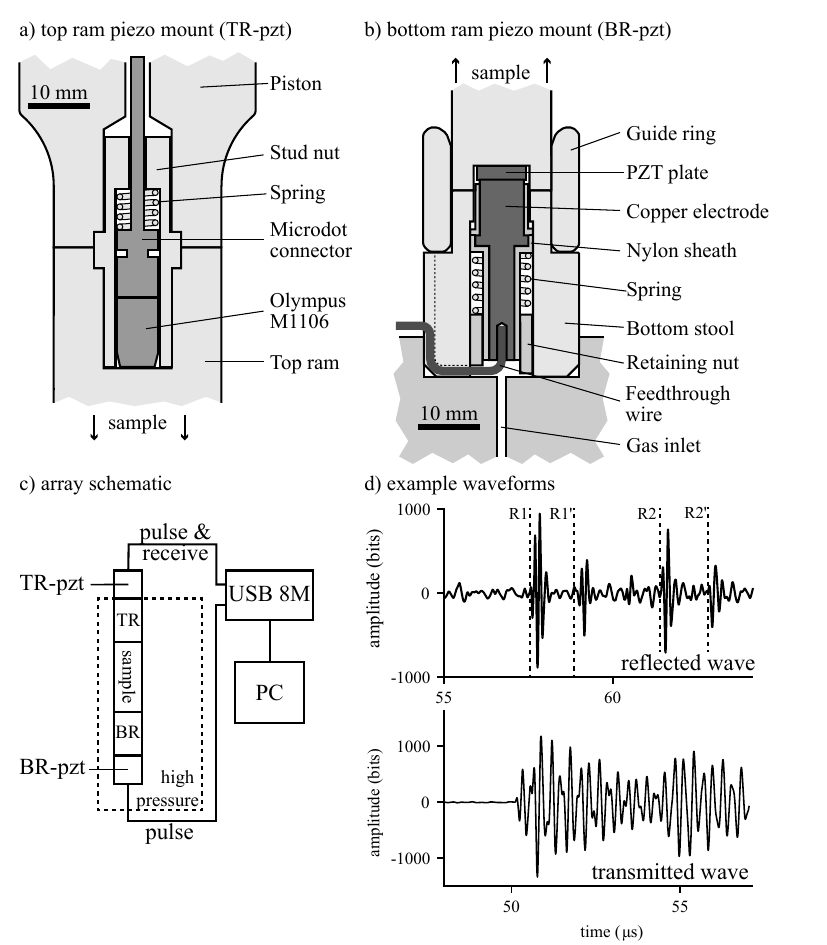}
\caption{\label{Fig:PZT_detail}Detail of piezo electric ceramic placement in a) the top ram, within a modified stud nut and b) glued into a recess at the base of the bottom ram. A spring loaded electrode mounted within a modified bottom stool is used to provide excitation to the bottom ram transducer. Inset c) shows a schematic illustration of the PZT array used to measure acoustic velocity. Inset d) shows examples of reflected and transmitted waveforms received by the top ram transducer. R1 is the reflection from the interface between the top ram and the sample spacer, R1$^\prime$ the focus of internal reflections generated by R1. R2 is the reflection from the interface between the sample spacer and the sample, and R2$^\prime$ corresponds to the focus of internal reflections generated by R2.}
\end{figure}

Our acoustic array consists of two Lead Zirconium Titanate ceramic (PZT-5H) transducers, positioned outside the hot zone, within the top and bottom ram respectively (figure \ref{Fig:PZT_detail}).
The top ram sensor is a commercially made acoustic transducer (Olympus M1106, central frequency 4.7 MHz) and is positioned against a flat recessed section at the end of the M12 threaded section within the top ram (figure \ref{Fig:PZT_detail}a).
A modified stud connector is used to spring load the transducer and align it with the ram axis.
The bottom ram sensor consists of a PZT plate positioned into a 4mm deep, 9 mm diameter recess at the base of the bottom ram, bonded using silver loaded epoxy (figure \ref{Fig:PZT_detail}b).
The bottom of the transducer contacts a spring-loaded copper electrode housed within the bottom stool which is connected to a feedthrough wire.
This arrangement of transducers results in easy maintenance of the system, allowing us to swap or remove transducers without removing the upper and lower closure nuts.
Both transducers are pulsed and monitored using an Eurosonic USB8M pulse-receiver unit (figure \ref{Fig:PZT_detail}c).
Although the current system is set up to make active measurements, a separate oscilloscope could be set-up to record passive signals.

\subsection{\label{sec:level2}Transmitted wave measurements}

Wave speed in the sample is measured using the pulse-transmission method \citep{Birch1960}.
To do this, the bottom ram PZT transducer is excited by a 2.5 MHz, 200 V pulse, which is received by the sensor positioned within the top ram (figure \ref{Fig:PZT_detail}c).
The receiving sensor is pre-amplified at 40 dB and recorded at 100 MHz with 12 bit accuracy, signal quality is improved by stacking 256 individual measurements at each time interval.
In this configuration the velocity can be measured every 0.5 s.

During experiments, relative changes in arrival times are computed by cross-correlation of individual waveforms against a manually picked "master" waveform.
In order to increase the arrival time accuracy, the raw waveforms are sub-sampled at 10 times the original sampling frequency using cubic interpolation.
Following re-sampling, the signals are cut with a window centered around the arrival time.
The arrival time is then computed by cross correlating the trimmed signals, with the offset time, $\delta t$ taken as the time difference corresponding to the maxima of the correlation function between the master and test waveform.

The travel time through the sample is calculated by removing the delay resulting from the sample assembly.
Time delay through the loading column changes with load, pressure and temperature; the column shortens elastically as a result of load and pressure, but it can also expand thermally.
To measure delays resulting from the sample assembly we performed a series of calibration runs using a flawless fused silica blank, which has a known wave velocity that does not vary significantly with pressure.
We cycled load on the sample at 200 and 400 MPa confining pressure, and also performed a cycle at a temperature of 200$^\circ$C.

\begin{figure}
\includegraphics{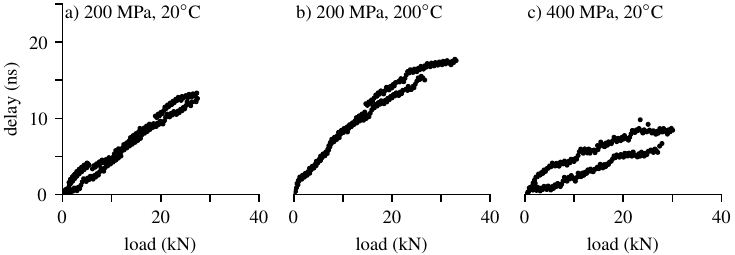}
\caption{\label{Fig:trans_delay}Time residual after removing contributions to time delay resulting from elastic shortening of the load column at a range of test conditions. Inset a) 200 MPa at room temperature, b) 200 MPa at 200 $^\circ$C and c) 400 MPa at room temperature.}
\end{figure}

When removing all possible contributions relating to the changes in sample and loading piston length we still observe a small time delay (0-20 ns) that varies with load (figure \ref{Fig:trans_delay}).
This delay is interpreted to arise from changes of the contact compliance of loaded interfaces \citep{Pyrak-Nolte1987,Mollhoff2009}.
Interface group time delay is directly proportional to contact compliance \citep{Pyrak-Nolte1987}.
Contact compliance is also known to decrease with increasing load \citep{Greenwood1967,Pohrt2012}, therefore the remaining time residual is interpreted to result from these changes.
\citet{Greenwood1967} predict that compliance is approximately proportional to the ratio of root mean square of surface elevation to normal force. 
All contacting surfaces are ground using a 5 $\mu$m diamond wheel to ensure parallelity prior to tests.
Therefore, at low compliance interfaces, such as the nominally flat ground metal and rock surfaces used here, the change in arrival time is expected to be linear with load \citep{Pyrak-Nolte1987}.
In calibrations with a fused silica blank, the remaining interface time delay demonstrates a reasonable degree of linearity up to 30 kN load at all test conditions (figure \ref{Fig:trans_delay}).

Since interface time delay depends on the composite elastic properties of contacting interface \citep{Schoenberg1980}, then the time delay will also vary with the sample used.
Therefore, we need to correct each experiment on an individual basis.
To do so we first correct $\delta t$ for load column and sample shortening.
Then we fit the initial change in $\delta t$ versus differential load immediately following the "hit-point" between 1 and 4 kN.
At such low loads any cracking in the sample will be preferentially aligned parallel with the sample axis, therefore axial P-wave speed is expected to remain nearly constant and changes in arrival time will be dominated by changes to the interface time delay. 
The obtained linear fit of $\delta t$ versus load is then used to correct $\delta t$ as a function of load during the entire experiment.

\begin{figure}
\includegraphics{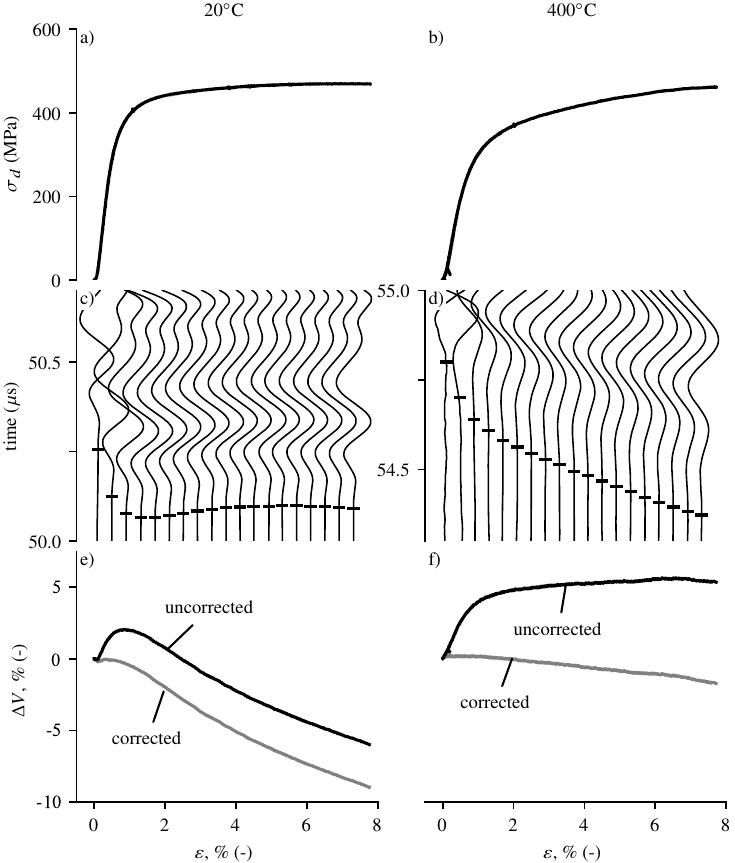}
\caption{\label{Fig:Transmission}Examples of mechanical data and transmitted wave measurements during the deformation of Solnhofen limestone samples at 20$^\circ$C and 400$^\circ$C and 200 MPa confining pressure.
Inset a) is the stress strain curve obtained at 20$^\circ$C and b) the stress strain curve at 400$^\circ$C. 
Insets c) and d) show example raw transmitted first arrival waveforms received at the top transducer normalised by the maximum possible value (= 2048 bits), the horizontal bars indicate the arrival time obtained using cross correlation.
Inset e) and f) show the computed compressional wave velocity changes. 
The uncorrected data (black curve) is calculated without any correction for interfacial load delay, and the corrected data (grey curve) shows the effects of using the interfacial load delay correction.}
\end{figure}

To test our system, we deformed two samples of the Solnhofen limestone at 20$^\circ$C and 400$^\circ$C, at a strain rate of $1\times10^{-5}$ s$^{-1}$ and a confining pressure of 200 MPa. 
At low strain, differential stress increases linearly with strain, corresponding to elastic loading of the sample (figure \ref{Fig:Transmission}a and b).
After the initial elastic loading, the differential stress rolls over with strain corresponding to yield of the sample.
Differential stress continues to increase after yield during the entire experiment, and indicates strain hardening behaviour.
Yield stress and final differential stress levels are higher in the room temperature experiment (figure \ref{Fig:Transmission}a), in comparison with the experiment at 400$^\circ$C (figure \ref{Fig:Transmission}).

The differences in mechanical data are also reflected in the measured arrival times (figure \ref{Fig:Transmission}c and d), and subsequent velocity computation (figure \ref{Fig:Transmission}e and f).
Both picked arrival times show a reduction in travel time during initial loading (figure \ref{Fig:Transmission}c and d), largely as a consequence of interface compliance changes and elastic shortening of the sample.
Following the initial decrease in arrival time, at 20$^\circ$C the arrival time increases slightly before remaining relatively constant (figure \ref{Fig:Transmission}c).
In the 400$^\circ$C experiment, the arrival time continues to decrease at smaller rate than the initial decrease (figure \ref{Fig:Transmission}d).

To test the correction, velocity was in calculated two ways; first by removing the known, load dependent travel time through the sample assembly (figure \ref{Fig:Transmission} e and f, "uncorrected" black curve), and then by removing the estimated interfacial time delay (figure \ref{Fig:Transmission}b and d, "corrected" grey curve), based on the arrival time changes during initial loading of the sample (differential load of 1-4 kN).
Velocity is simply the ratio of sample length to travel time, and is normalised by the initial velocity at the experiment hit point.
The corrected sample P wave velocity drops considerably after initial elastic loading at 20$^\circ$C (figure \ref{Fig:Transmission}e), but remains relatively constant at 400$^\circ$C (figure \ref{Fig:Transmission}f).
The correction removes the large increase in velocity during the initial elastic loading of the sample (figure \ref{Fig:Transmission} e and f grey curves).
However, the correction procedure maintains quantitative differences between individual experiments, and does not significantly affect relative changes after yield.

Changes in P wave speed can be used to infer the microstructural state of the sample.
Velocity reduces when tensile microcracks propagate in the sample.
At 20$^\circ$C the large decrease in wavespeed suggests that cracking is the dominant deformation mechanism (figure \ref{Fig:Transmission}e), which is consistent with the observation of \citet{Baud2000}. 
At 400$^\circ$C however, the relatively small decrease in wavespeed results from the increased activity of plastic deformation mechanisms which are expected at these conditions \citep{Heard1960}.
Plastic deformation suppresses crack growth, which can explain the small decreases in wavespeed.

The key advantage of having access to wave velocity \textit{in situ} is clear when comparing the relatively minor differences in stress-strain behaviour between the tests conducted at 20$^\circ$C and 400$^\circ$C (figures \ref{Fig:Transmission}a, b) to the large qualitative difference in wave velocity evolution (figures \ref{Fig:Transmission}e, f). 
Stress-strain behaviour cannot be used here to distinguish between microscale deformation processes, and careful microstructural work on post-mortem samples would be needed to establish those, with caveats due to potential damage and cracking during quenching and decompression. 
\textit{In-situ} wave velocity measurements largely circumvent those issues, at least to identify regimes where microcracking dominates.

\subsection{\label{sec:level2}Reflected wave delay as a piezometer}
The most accurate and reliable measurements of stress on samples in a pressure vessel are obtained from an internal load cell; located within the high pressure chamber where seal friction does not contribute to the load.
Interfacial reflected wave time delay is also sensitive to changes in interface compliance \citep{Schoenberg1980,Pyrak-Nolte1987}, and could therefore be used to measure load internally.  
Here we present a new technique to obtain load using interfacial group delay, which allows differential stress to be measured accurately without needing to correct for seal friction.

\begin{figure*}
\includegraphics{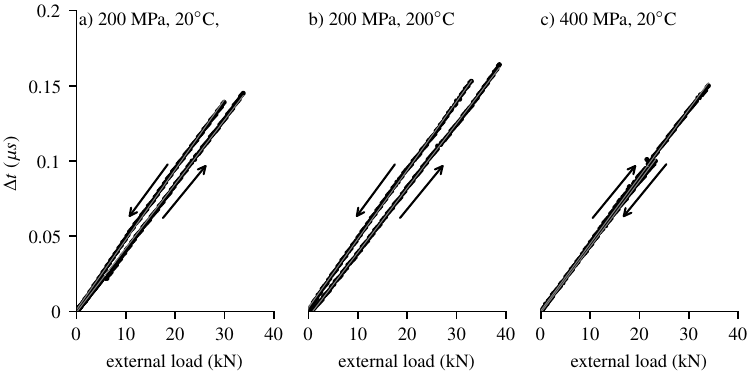}
\caption{\label{Fig:reflection1}Top piston reflection delay calibration during load cycles with a fused silica blank at a range conditions with hysteresis due to seal friction removed. Inset a) load cycle at 20$^\circ$C, 200 MPa confining pressure, b) 200$^\circ$C, 200 MPa confining pressure and c) 20$^\circ$C, 400 MPa confining pressure. Arrows indicate loading sense, black scatter is the raw data and the overlaid grey curve a linear best fit of the data.} 
\end{figure*}

\begin{figure*}
\includegraphics{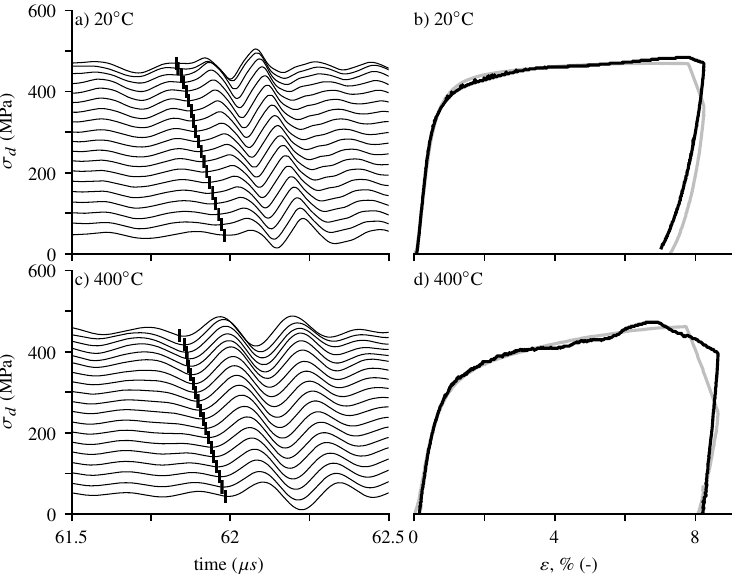}
\caption{\label{Fig:reflection2}Top piston reflection delay calibration from experiments conducted with the Solnhofen limestone at 20$^\circ$C and 400$^\circ$C and 200 MPa confining pressure. Insets a) and b) show raw waveforms obtained during loading of the sample which is used to derive the calibration between load and delay time. Vertical bars indicate arrival times obtained using cross correlation. Insets c) and d) show the use of this calibration to obtain stress. The stress strain curves in grey are the data obtained from the external load cell, and the black curves data obtained from the reflected wave delay. Fluctuations in stress after yield in the high temperature experiment are a result of temperature changes resulting from argon convection.} 
\end{figure*}

In order to test reflected wave delay as a proxy for differential stress, we recorded reflected waves during load cycles with a fused silica blank (figure \ref{Fig:reflection1}). 
The 10 mm diameter silica sample was load cycled at 20$^\circ$C and 200$^\circ$C at 200 MPa confining pressure, and also at 20$^\circ$C at 400 MPa.
We then used the second reflected arrival (R2, see figure \ref{Fig:PZT_detail}d), corresponding to the sample spacer-sample interface, and used cross correlation to obtain the change in arrival time (figure \ref{Fig:PZT_detail} a and c).
In these load cycles a delay of $\approx$4 ns kN$^{-1}$ is typical (figure \ref{Fig:reflection2}a and c), which when combined with a subsampling frequency of 1 GHz, results in a resolution of about 2 MPa differential stress.

We now apply this technique to the previously discussed experiments performed with the Solnhofen limestone, where we also measured reflected wave arrivals during the test.
The sample spacer-sample interface reflection was used in the cross correlation analysis (R2, figure \ref{Fig:PZT_detail}d).
The calibration was made on an individual basis for each experiment, using the measured changes of $\delta t$ during elastic loading phase of the experiment.
Results show that this measurement is able to reproduce the stress measurements accurately at 20$^\circ$C (figure \ref{Fig:reflection2} b), and reveals extra details during the unloading phase of the experiment.
In particular, it provides an accurate measurement of the stress-strain behaviour at the onset of unloading and throughout the unloading phase.
The inflection in stress-stress behaviour observed at low load (figure \ref{Fig:reflection2}b and d) can provide information about residual stresses stored within the sample \citep[e.g.][]{Hansen2019}.
These observations are normally obscured in external load measurements due to hysteresis resulting from the top plug seal friction.
At higher temperatures, the large stress changes during initial loading and unloading are well matched.
However, during strain hardening, temperature fluctuations lead to variations in the stress computed using this measurement (figure \ref{Fig:reflection2}d).

Looking to other applications, reflected wave measurements could also be used in stress relaxation experiments where piston motion is arrested.
In these tests, the differential stress state is obscured by seal friction, which this technique overcomes.
This technique also has potential for use in other apparatus, where space is limited, or it is not possible to position a sensor due to harsh environmental factors.
In particular, it may be a useful technique to obtain load in solid medium apparatus which suffer from large uncertainties resulting from seal friction \citep{Holyoke2010}.
It would also directly complement the method suggested by \cite{Moarefvand2021}, where arrival times of use transmitted waves are used to infer column length in a Griggs apparatus. 

\section{\label{sec:level1}Conclusions}

In this manuscript we discussed and detailed the modifications, complete refurbishment and upgrades made to a high-pressure high-temperature gas apparatus in order to bring it back into working service.
The apparatus can actively monitor acoustic wavespeed during deformation up to confining pressures of up to 1000 MPa and temperatures of up to 700$^\circ$C.
We discussed the construction and calibration of an internal furnace, and the effects of varying pressure and temperature on the temperature profile, showing that the hot zone elevation decreases with increasing pressure.
In calibrating the acoustic velocity measurements we observed a load dependant time residual which is not accounted for by elastic shortening of the load column.
The load dependant time residual is interpreted to result from changes in interface compliance which result in an interfacial time delay which must be subtracted from the travel time measurements.
The interface delay is material dependant, and we therefore present a technique to estimate this on an individual experimental basis, and show how it influences results.
As this delay is unique function of load we demonstrate how it can be used to measure load internally, using reflected pulses, which gives a more accurate measurement of sample stress.
Our new technique for measuring load could be developed further to measure load accurately in other similarly inaccessible environments, where it is not possible to position a conventional load cell.
Our acoustic array could easily be upgraded to make passive measurements of acoustic emission from the sample.
S-wave transducers could also be used to measure the shear wave velocity of the sample during deformation.

\begin{acknowledgments}
We wish to acknowledge John Bowles, Steve Boon and Neil Hughes for technical support, as well as Paul and Mark Freeman of Sovereign pressure products for support with the gas intensification pump.
Colin Jones and Phil Meredith provided highly valuable historical records about the apparatus.
Experimental techniques were enriched by discussions with Toshi Shimamoto, Ian Jackson, Kathryn Hayward, Lars Hansen, Greg Hirth, Dave Dobson, Julian Mecklenburgh and Alexandre Schubnel.

This study has received funding from the European Research Council (ERC) under the European Union’s Horizon 2020 research and innovation programme (Grant agreement No. 804685) to NB  and from the UK Natural Environment Research Council (Grant agreement NE/M016471/1 to NB and TMM).
\end{acknowledgments}

\section*{Data Availability Statement}
The data that supports the findings of this study are available within the article and its supplementary material.

\bibliography{/Users/christopherharbord/Documents/library.bib}

\end{document}